\documentclass[10pt,superscriptaddress,twocolumn,amsmath,amssymb,aps,prx]{revtex4}
\usepackage{times}
\usepackage{amsfonts}
\usepackage{mathrsfs}
\usepackage{graphicx}
\usepackage{dcolumn}
\usepackage{bm}
\usepackage{color}

\usepackage[colorlinks,bookmarks=false,citecolor=blue,linkcolor=red,urlcolor=blue]{hyperref}

\def\be{\begin{equation}}       \def\ee{\end{equation}}
\def\bea{\begin{eqnarray}}      \def\eea{\end{eqnarray}}

\begin{document}
\title{ Unconventional High Temperature Superconductivity in Cubic Zinc-blende Transition Metal Compounds }

\author{Qiang Zhang}

\affiliation{Beijing National Laboratory for Condensed Matter Physics,
and Institute of Physics, Chinese Academy of Sciences, Beijing 100190, China}

\author{Kun Jiang}
\affiliation{Beijing National Laboratory for Condensed Matter Physics,
and Institute of Physics, Chinese Academy of Sciences, Beijing 100190, China}
\affiliation{Department of Physics, Boston College, Chestnut Hill, MA 02467, USA}

\author{Yuhao Gu}
\affiliation{Beijing National Laboratory for Condensed Matter Physics,
and Institute of Physics, Chinese Academy of Sciences, Beijing 100190, China}
\affiliation{Beijing National Laboratory for Molecular Sciences, State Key Laboratory of Rare Earth Materials Chemistry and Applications, Institute of Theoretical and Computational Chemistry, College of Chemistry and Molecular Engineering, Peking University, 100871 Beijing, China}

\author{Jiangping Hu}\email{jphu@iphy.ac.cn}
\affiliation{Beijing National Laboratory for Condensed Matter Physics,
and Institute of Physics, Chinese Academy of Sciences, Beijing 100190, China}
\affiliation{Kavli Institute of Theoretical Sciences, University of Chinese Academy of Sciences,
Beijing, 100190, China}

\begin{abstract}
We consider possible high temperature superconductivity (high-$T_c$) in transition metal compounds with a cubic zinc-blende lattice structure. When the electron filling configuration in the d-shell  is close to $d^7$, all three $T_{2g}$ orbitals are near half filling with strong nearest neighbor antiferromagnetic (AFM) superexchange interactions. We argue that upon doping, this electronic environment can be one of ``genes" to host unconventional high $T_c$ with a time reversal symmetry broken $d_{2z^2-x^2-y^2} \pm i d_{x^2-y^2}$ pairing symmetry. With gapless nodal points along the diagonal directions, this  state is a direct three-dimensional analogue to the two-dimensional $B_{1g}$ d-wave state in cuprates. We suggest that  such a case may be realized in electron doped CoN, such as CoN$_{1-x}$O$_x$ and (H, Li)$_{1-x}$CoN.  
\end{abstract}

\pacs{74.10.+v, 74.20.Mn, 74.20.Rp, 71.10.Li}

\maketitle

The superconducting mechanism of unconventional high-$T_c$ in cuprates\cite{Bednorz1986} and iron-based superconductors\cite{Kamihara2008-jacs}  remains one of the most challenging problems in condensed matter physics\cite{zaanen-np06towards,Norman-sci11,leggett-np06}.  A correct answer to this problem  should  be able to  guide us to identify or predict new materials with potential high-$T_c$. 

Recently, we have identified that a key character, called  the electronic gene, which separates these two classes of high-$T_c$ materials from other transitional metal compounds, is that the d-orbitals with the strongest in-plane d-p couplings in both high-$T_c$ families are  isolated near Fermi  energy\cite{Hu-tbp,Hu-genes, Hu-s-wave}.   The gene can only be realized by a specific collaboration through cation-anion  complexes,  global lattice structures, and  specific electron filling configurations in the d-shell of transition metal atoms. In  cuprates,  the  d$_{x^2-y^2}$  e$_g$ orbital  is  isolated near Fermi energy in  a  two-dimensional (2D) Cu-O square lattice formed by corner-shared CuO$_6$  octahedra (or CuO$_4$ square planar)  in a $d^9$ filling configuration of  Cu$^{2+}$.  In iron-based superconductors,  there are two $T_{2g}$ d$_{xy}$ -type orbitals which are isolated near Fermi energy with the $d^6$ filling configuration of Fe$^{2+}$ in a Fe(Se/As)  2D square lattice formed by edge-shared Fe(Se/As)$_4$ tetrahedra\cite{Hu-tbp,Hu-genes,Hu2012-s4}.

In order to justify the above idea, we must discover new families of high-$T_c$.  We have suggested to  realize the gene condition in  the  $d^7$  and $d^8$ filling configurations.  Up to now, we have predicted that the $d^7$ gene  condition can be realized   in a 2D hexagonal  layer formed by edge-shared trigonal bipyramidal complexes\cite{Hu-tbp} or in a 2D square lattice formed by the corner-shared tetrahedra\cite{Hu2017ASuperconductors, CuInCoTe,HU-sb18},  and the $d^8$ gene condition exists in a 2D square lattice formed by Ni-based mix-anion octahedra\cite{Le2017ASuperconductors}.  Unfortunately, all these proposals have not been materialized. 

To realize the gene condition, it is easy to notice that  a quasi-2D layer is generally required  because of the nature of the d-orbital spacial  configuration.  In fact, all above examples are quasi-2D. However, there may be one exception.  In the $d^7$ filling configuration with corner shared tetrahedra, all three $T_{2g}$ orbitals can participate in superconducting pairing.  These three orbitals together form a three-dimensional (3D) irreducible representation in a cubic lattice structure. Therefore, even in a 3D cubic lattice structure, they can be fully isolated near Fermi energy. 

\begin{figure}[ht]
    \centering
    \includegraphics[width=.45\textwidth]{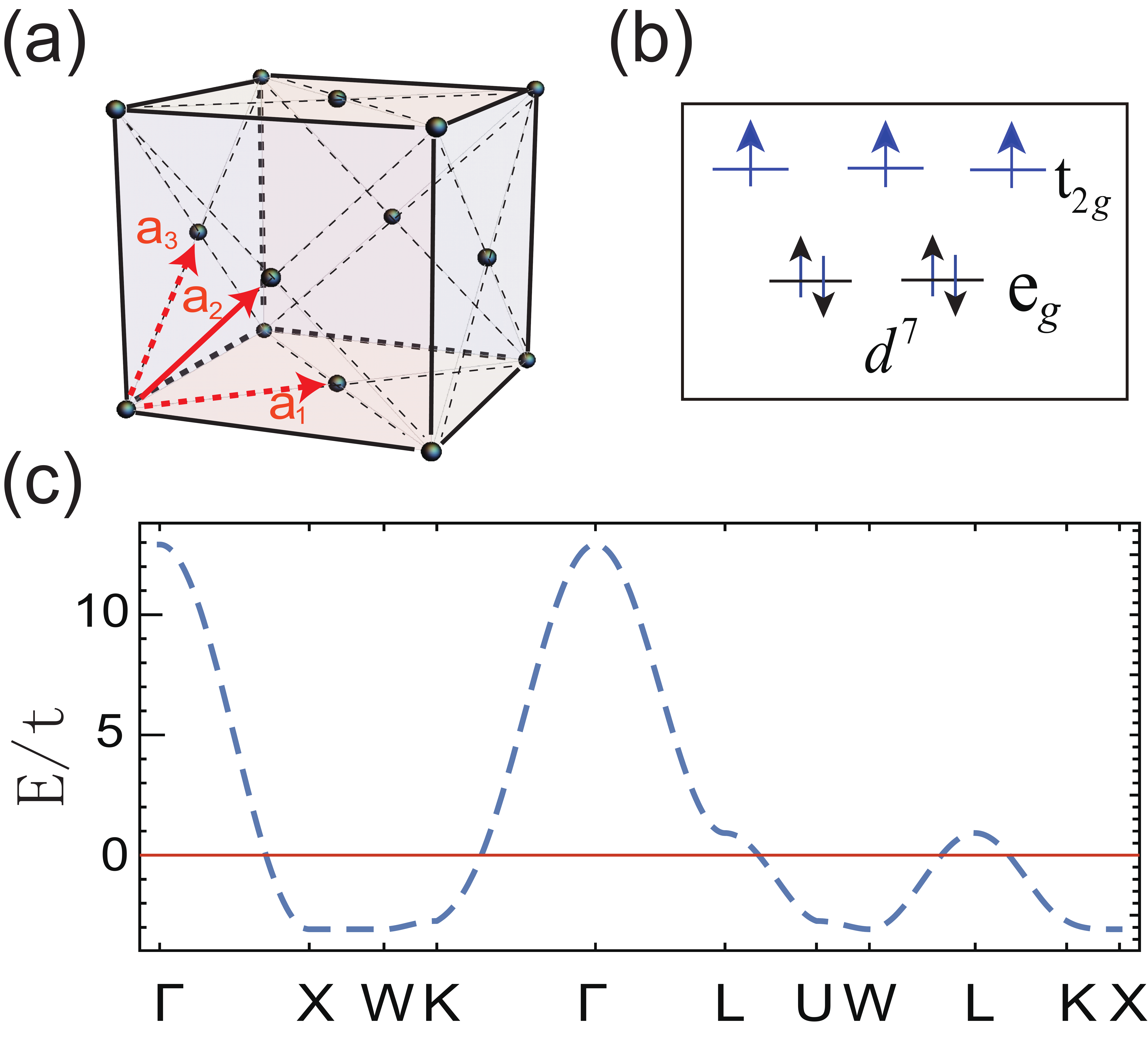}
    \caption{(a) The face center cubic structure with sketched lattice vectors. (b) The tetrahedra crystal field splitting of the five $3d$ orbitals. (e) The energy dispersion of the single band model along high symmetry lines.  }
    \label{fig:fcc}
\end{figure}

The  zinc-blende lattice structure which is formed by corner shared tetrahedra is an ideal lattice structure to test this potential exception. Different from the previous examples and proposals, the lattice of the shared tetrahedra in the zinc-blende structure is essentially 3D. Locally, the tetrahedra symmetry is fully respected.  The three $T_{2g}$ orbitals are degenerate and have kinematics  in all three spacial directions. If we consider a transition metal compound with  the zinc-blende structure, the transition metal atoms form a face center cubic (FCC) lattice as shown in Fig.\ref{fig:fcc}(a). Antiferromagnetic superexchange couplings can be generated through anions among all three $T_{2g}$ d-orbitals due to the crystal field energy splitting in a tetrahedron complex as shown in Fig.\ref{fig:fcc}(b). Thus, in the first-order approximation, such an electronic environment can be described by a 3D  t-J model.

In this paper,  we study the possible superconducting state  of a t-J model in a 3D FCC lattice. Both one-orbital and three-orbital models are considered. In both cases, we find that   a   $d_{2z^2-x^2-y^2}\pm id_{x^2-y^2}$ pairing superconducting state, which breaks the time reversal symmetry,  is  the most favored near half filling upon hole doping. This $d\pm id$  state is a spacial 3D analogue to the 2D $B_{1g}$ d-wave state in cuprates. Similarly, the gap function vanishes  along the diagonal lines, resulting in nodal points  in the superconducting states if  Fermi surfaces insect with the lines.    The extended s-wave state is the least favored followed by  chiral t-wave states. We suggest that  the case can be potentially realized in electron doped CoN\cite{CoN95}, such as CoN$_{1-x}$O$_x$\cite{CoNO04,CoNO15} and (H, Li)$_{1-x}$CoN\cite{reddy2013li,MPLB18-1rsprinciple}.

We first consider a one-orbital  t-J model in a FCC lattice.  We assume that the effective hopping between the  d-orbitals is mainly  through d-p couplings and is dominated by the nearest neighbor(NN) hopping.  As we will show later, if we ignore the inter-orbital hopping, the kinematics of  each $T_{2g}$ orbital in the zinc-blende lattice is close to that of an isotropic s-orbital. Therefore, for simplicity, we start with an isotropic single-orbital model. By setting the second NN distance as one, the kinetic part of the Hamiltonian can be written as $H_t=\sum_{k\sigma} a^\dagger_{k\sigma} H_k a_{k\sigma}$ with
\begin{align}\label{equ:1tb}
    H_k=4t( c_xc_y+c_zc_y+c_yc_z)-\mu,
\end{align}
where  $c_{\alpha}\equiv\cos{(\frac{k_\alpha}{2})}$, $\alpha=x,y,z$ and $\sigma$ is the spin index. Here $t$ is the nearest neighbor hopping amplitude and $\mu$ is the chemical potential, with $\frac{\mu}{t}=-0.92$ at half filling. The band structure along the high symmetry momenta is shown in Fig.\ref{fig:fcc}(c).  Assuming the NN AFM superexchange coupling $J$, the total Hamiltonian  for a standard t-J model can be written as
\begin{align}
    H= \hat P (H_t)+J\sum_{<ij>}(S_iS_j-\frac{1}{4}n_in_j),
    \label{tj}
\end{align}
where $\hat P$ is the projection operator to forbid double occupancy at a single site and $S_i(n_i)$ denotes the spin (density) at site-$i$.

In a mean field solution, the above Hamiltonian in 2D is known to result in a phase diagram between a long range AFM state and a superconducting state\cite{Liu-Kotliar-prb88},   which was known to be qualitatively consistent with experimental results in cuprates\cite{Liu-Kotliar-prb88, PatrickLee01renormalized,Patrick-RMP06} and iron-based superconductors\cite{Seo-prl08pairing,Qimiaosi_slavespin,Qimiaosi09-correelation}.  There is no much difference for this result in 3D space\cite{Ichinose-prb01}. Moreover, without doping, the antiferromagnetism and magnetic frustration in the FCC structure have already been extensively studied in literature\cite{PR52-molecularfield,PR58-fccMag,SUN18-fccmag,PR65-fcc}. Therefore, we ignore the AFM phase and solely focus on the superconducting state of the above model upon doping in this paper.

Upon doping, the superconducting state of the above 3D model  is more interesting because the lattice symmetry is much larger in three dimensions than in two dimensions.  The FCC  lattice is governed by the $O_h$ point group. The superconducting pairing functions belong to the irreducible representations of the $O_h$ group. As the superconducting pairing is induced by the NN AFM interaction, the superconducting order must be in the spin-singlet pairing channel and  carry specific momentum form factors.   Considering the equal pairing strength for all  NN bonds,  we have
\begin{eqnarray}
  & &   \Delta^{A} =4\delta_s (c_xc_y+c_yc_z+c_xc_z), \label{equ:pairingA}\\
  & &  \Delta^{E_{\pm}}  = 2\delta_d(2c_xc_y-c_xc_z-c_yc_z\pm i\sqrt{3}(c_yc_z-c_xc_z)), \\
   & & \Delta^{T} = -4\delta_t (s_xs_y+e^{i\theta_1}s_ys_z+e^{i\theta_2}s_xs_z),\label{equ:pairingT}
\end{eqnarray}
where $s_\alpha=sin(\frac{k_\alpha}{2})$.  $ \Delta^{A} $ represents the extended s-wave,    $\Delta^{E_ {\pm}}$ represents a time reversal symmetry broken superconducting state $d_{2z^2-x^2-y^2}\pm i d_{x^2-y^2}$ that belongs to a 2D $E_g$ irreducible representation. It can also be written  symmetrically as $4\delta_d(c_xc_y+e^{\pm i2\pi/3} c_xc_z+e^{\pm i4\pi/3} c_yc_z)$. Obviously, this $d\pm id$ gap function vanishes  along the diagonal lines $|k_x|=|k_y|=|k_z|$, as a 3D analogue of the 2D $B_{1g}$ d-wave for the cuprates. However, in this one-orbital model, near half filling upon hole doping, there is no Fermi surface along the $\Gamma$-$L$ lines as depicted in Fig.\ref{fig:fcc}(c). Therefore the state is still fully gapped.  $ \Delta^{T}$ represents a t-wave superconducting state that belongs to a 3D $T_{2g}$ 
irreducible representation.  In the $\Delta^{T}$ state,  besides the pairing strength $\delta_t$, we have two additional phase parameters $\theta_{1,2}$.  From the standard Ginzburg-Landau theory analysis\cite{Sigrist-RMP91}(see the App.\ref{app:landau}), we have two types of  t-wave phases, the time reversal invariant (TRI) and chiral phases. For the TRI phase,  we can take $\theta_1=\theta_2=0$ and for the chiral one,  we can take $\theta_1=2\theta_2=4\pi/3$. 
\begin{figure}[ht]
    \centering
    \includegraphics[width=.45\textwidth]{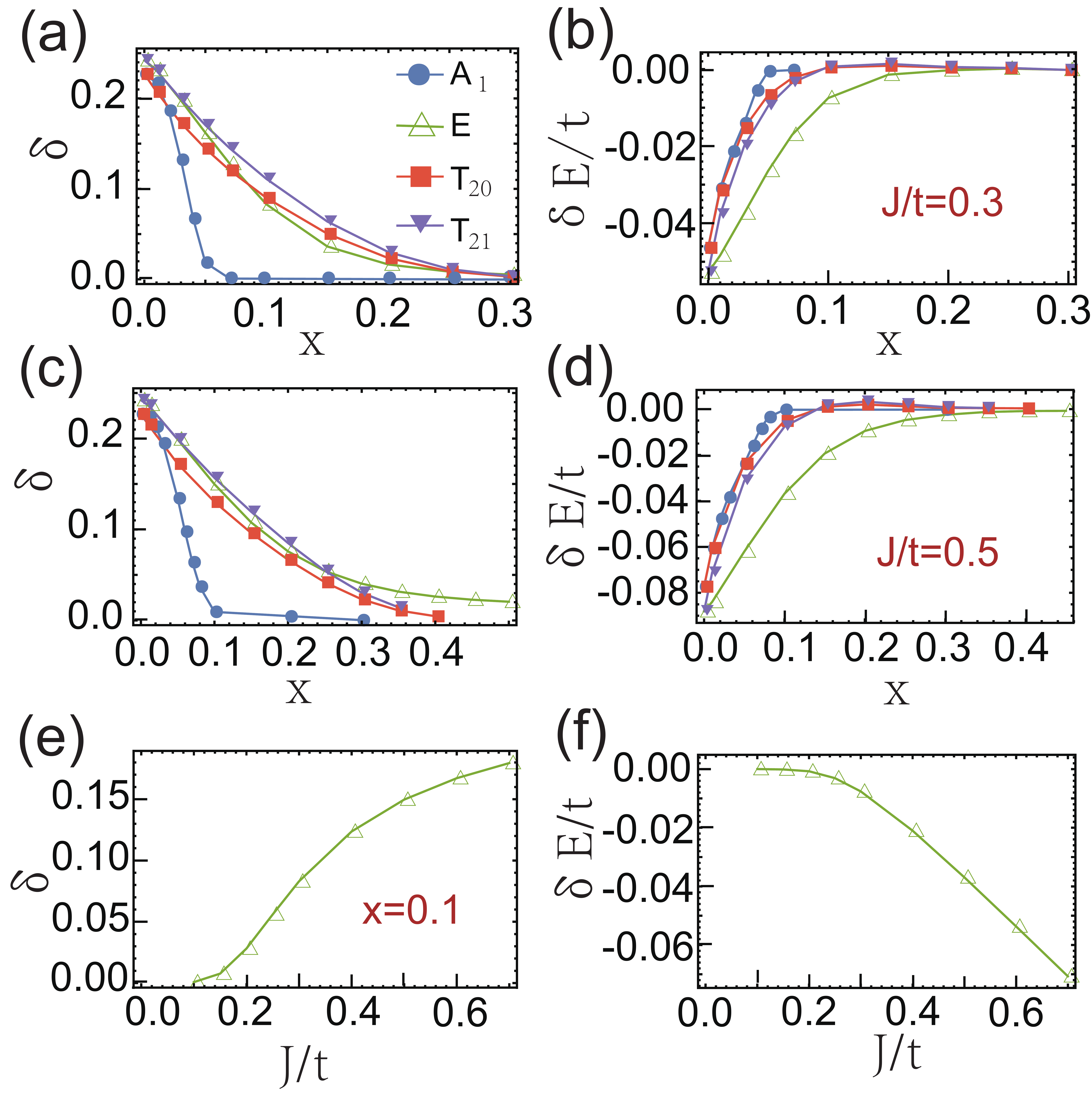}
    \caption{ The results from the slave boson mean field solution of  the single-orbital t-J model: (a,c) the pairing strength $\delta_{s/d/t}$  and (b, d) the ground state energy   $\delta E$  with respect to the normal state   versus \textit{hole} doping at J/t=(0.3, 0.5); here $T_{20}$ and $T_{21}$ denote the TRI and chiral $t$-wave pairing states, respectively. (e, f) the order parameter $\delta_d$ and the ground state energy as a function of J/t  with a hole doping level $x=0.1$.  }
    \label{fig:1bd}
\end{figure}

We use the slave boson mean-field approach\cite{Sigrist05slave,QHWang-prb04,PatrickLee01renormalized,gu-prb19Ni} to calculate the  superconducting states in the hole doped region. The results are reported in Fig.\ref{fig:1bd} with $J/t=0.3$ in (a,b) and $J/t=0.5$ in (c,d), respectively. 
The $E_g$ state wins over other two states.  As shown in Fig.\ref{fig:1bd}(a,c), the extended $s$-wave quickly diminishes with doping while  the superconducting orders in the $d\pm id$ wave $E_g$ state and  the  t-wave states appear to be comparable. However, as shown Fig.\ref{fig:1bd}(b,d), the $E_g$ state has much lower energy  than the t-wave states in relevant doping regions. We also notice that the time reversal symmetry broken phases  generally gain more condensation energy than the unbroken counterparts. For example, the chiral t-wave state has lower energy than the TRI  t-wave state. In Fig.\ref{fig:1bd}(e,f), we report the order parameter of the $d\pm id$ wave state and its condensed energy as the function of the AFM exchange coupling at a fixed doping $x=0.1$. 

This result can be easily understood by the Hu-Ding principle\cite{hu_local_2012}, which states the favored pairing symmetry is determined by the overlap strength between the momentum form factors of pairing functions and Fermi surfaces. For the extended s-wave, the d-wave and the t-wave states,  the form factors  peak    at $\Gamma$, $X$ and $L$ high symmetry points, respectively. Moreover, the d-wave can open bigger gaps near $X$, $W$ and $K$ points.  Thus from the energy dispersion in Fig.\ref{fig:fcc}(c),  the d-wave state wins near half filling upon hole doping.  Yet, with heavy electron doping to reach the $L$ (saddle) points, the t-wave can eventually become favored.

\begin{figure}[ht]
    \centering
    \includegraphics[width=.45\textwidth]{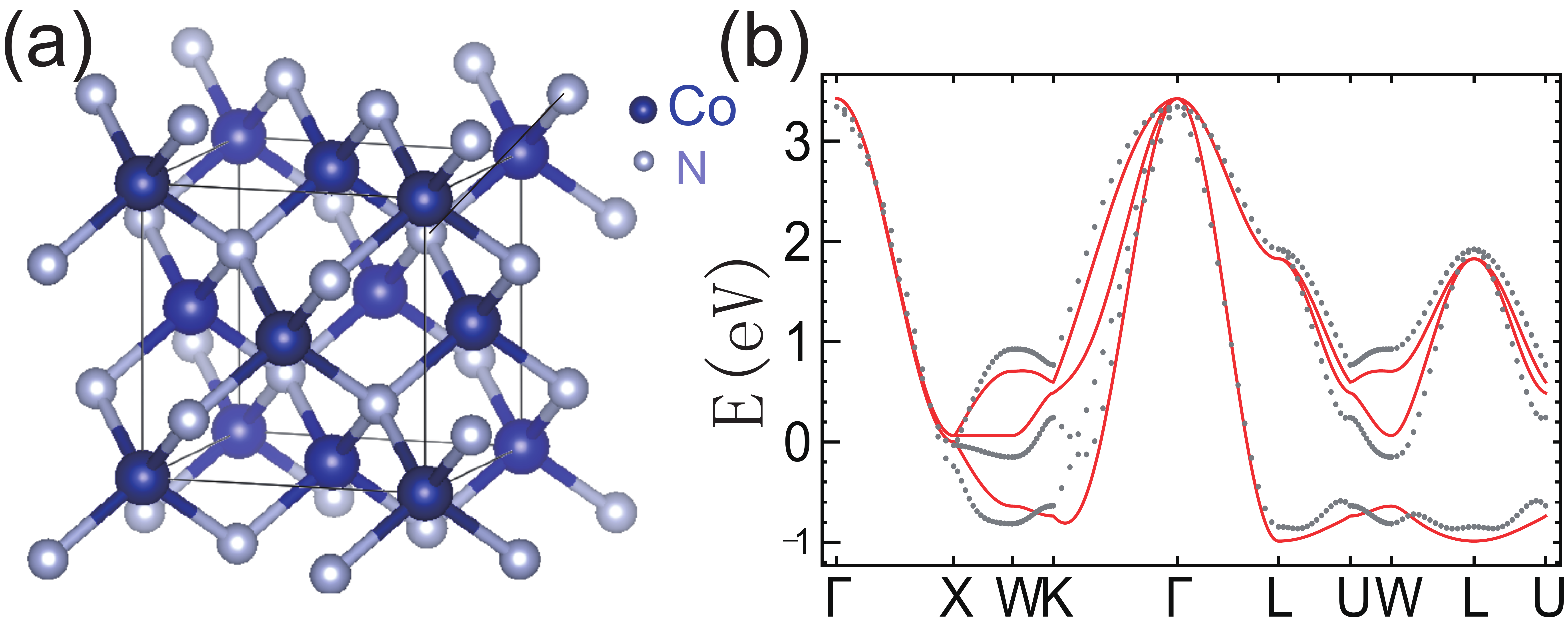}
    \caption{The zinc-blende structure for CoN (a) and its three-band dispersion along the high symmetry momenta (b). The NN bond hopping tight-binding dispersion  (solid red)  with $(t_1,t_2,t_3,t_4,\mu)=(0.214,0.206,0.235,0.084,-0.889)eV$ fits fairly well with the DFT result (dotted gray).}
    \label{fig:zb_disp}
\end{figure}

Now we consider all three $T_{2g}$ orbitals. In this case, it is important to note that   due to the lack of inversion symmetry,  zinc-blende structure belongs to the space group $F\Bar{4}3m (No.216)$ and its point group is $T_d$ instead of the $O_h$ group,  as depicted in Fig.\ref{fig:zb_disp}(a).  This symmetry difference allows additional hopping terms.  

Taking the cobalt nitrogen (CoN\cite{CoN95}) as an example which has the zinc-blende structure with a lattice constant $4.27\ \AA$ as depicted in Fig.\ref{fig:zb_disp}(a), its band structure  from density function theory (DFT\cite {Kresse1996,kresse1999,perdew1996}) is shown  in Fig.\ref{fig:zb_disp}(b) with the dotted gray line \footnote{ Our DFT calculations employ the VASP code\cite{Kresse1996} with the PAW method\cite{kresse1999} and the PBE\cite{perdew1996} exchange-correlation functional. The kinetic energy cutoff  of 600 eV for the planewaves, $\Gamma$-centered k-mesh of $16\times16\times16$ and energy convergence criterion of $10^{-6}$ eV are adopted. }.   The half-filled $T_{2g}$ orbital is achieved if we dope one more electron per Co atom into this material. 

 We simplify the band structure in a tight binding model with only the NN hopping among three orbitals. We find that the model  is a very good approximation to describe the band dispersion as indicated by the red solid lines in Fig.\ref{fig:zb_disp}(b). To be specific, we define the hopping matrix along $\mathbf{a}_1=(1,0,1)/2$ directions $T(\mathbf{a}_1)$ in the basis $\psi^\dagger(\mathbf{r})=(d^\dagger_{yz},d^\dagger_{xz},d^\dagger_{xy})$.  Due to the time reversal symmetry,  the mirror symmetry on the plane $(\Bar{1}10)$ and the two-fold rotation symmetry along $z$-axis, there are only   four free hopping parameters:
\begin{align}
    T(\mathbf{a}_1)=\left(\begin{array}{ccc}
       t_1      &t_3    &t_4    \\
       t_3      &t_1    &t_4    \\
      -t_4      &-t_4  &t_2
    \end{array}\right),
\end{align}
with $t_{1,2,3,4}$ being real numbers. The hopping matrix along other directions can be obtained by symmetry operations: $T(R_g\mathbf{a}_1)=R_gT(\mathbf{a}_1)R^{-1}_g$.   The three-orbital tight-binding Hamiltonian matrix can be written as
\begin{align}\label{equ:CoNtb}
    H_{3t}(\mathbf{k})=\sum_{\langle g \rangle} T(R_g\mathbf{a}_1)e^{i \mathbf{k}\cdot R_g\mathbf{a}_1}-\mu,
\end{align}{}
here $\langle g\rangle\in T_d $ ensures that $R_g \mathbf{a}_1$ runs over all the 12 NN vectors. Specifically, the matrix elements in $H_{3t}(\mathbf{k})$ are 
\begin{align}
    H_{3t,11}(\mathbf{k})=&4\big(t_1 (c_xc_y+c_xc_z)+t_2c_yc_z\big)-\mu,\nonumber\\
    H_{3t,12}(\mathbf{k})=&4\big(it_4(c_x-c_y)s_z-t_3s_xs_y\big),
\end{align}
and other terms can be obtained by cyclical permutation of the suffixes $x,y,z$. For CoN, we find that  a set of the parameters with $(t_1,t_2,t_3,t_4,\mu)=(0.214,0.206,0.235,0.084,-0.889)eV$  fits reasonably well to the DFT calculations as shown  Fig.\ref{fig:zb_disp}(b).   
We notice that $t_1\approx t_2$, which indicates that the band structure of the three-orbital model is similar to the single-orbital model if the interorbital hoppings  are ignored.  The interorbital hoppings only cause band splitting around $L$ and $W$ points. The $L$ points  splitting stems from the interorbital hopping $t_3$ term  and the $W$ points splitting comes from the  $t_4$ term. The small $t_4$ term stems from the absence of the inversion symmetry due to the existence of nitrogen atoms (see the App.\ref{app:symmetry}).  The above band structure is rather qualitatively generic to $T_{2g}$ d orbitals in the zinc-blende lattice structure. One can easily check  that with the effective hoppings between the d-orbitals being mainly induced by the d-p couplings, the lattice symmetry qualitatively provides such a band structure. 
 
By adding the AFM exchange interactions, we replace the kinetic energy term in  Eq.\ref{tj} by $H_{3t}$ to consider a multi-orbital t-J model\cite{kun-prl18,gu-prb19Ni} and study the possible superconducting states. As we only consider the spin-singlet pairing,  the representations of $T_d$ group are the same to the even-parity representations of $O_h$. Thus, the pairing symmetry analysis for the single-orbital model in FCC structure is applicable to the three-orbital model.   

\begin{figure}[ht]
    \centering
    \includegraphics[width=.45\textwidth]{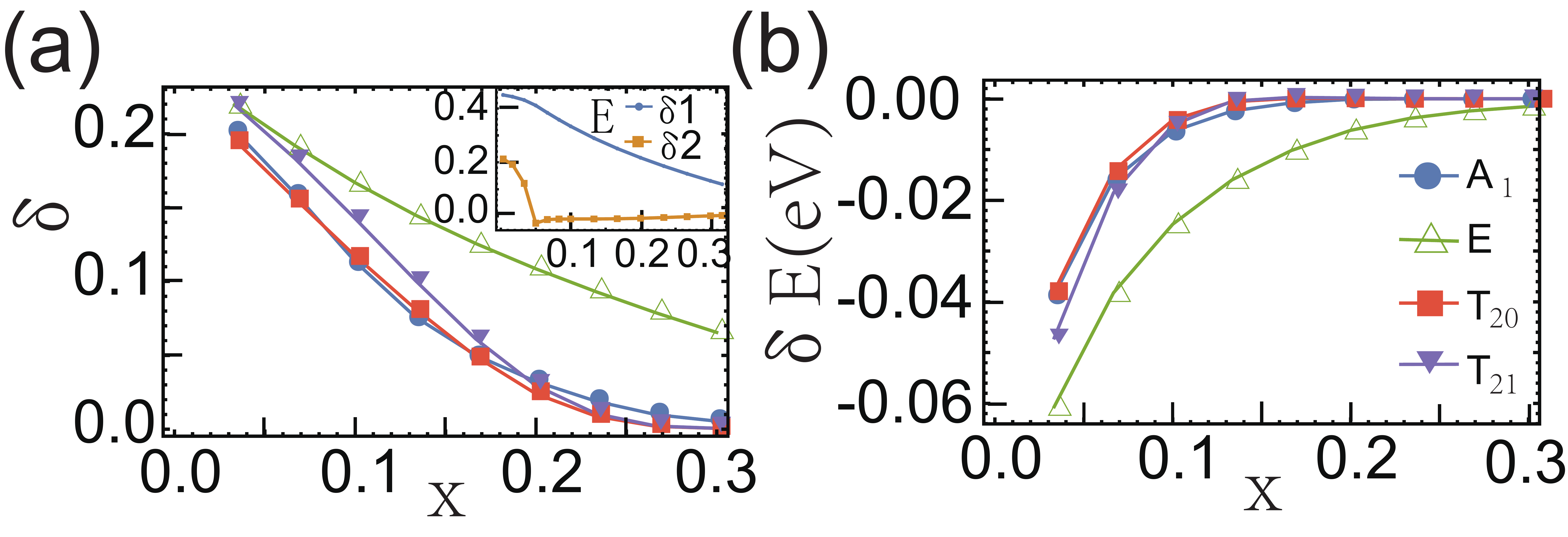}
    \caption{ The slave boson meanfield results for the three-orbital t-J model: (a) the pairing strength $\delta_{s/d/t}$ and (b) the ground state energy versus \textit{hole} doping away from half-filling.  Here the tight-binding parameters for CoN at Fig.\ref{fig:zb_disp} are used and $J=0.2$ eV. The inset in (a) depicts the intraorbital isotropic $(\delta_1)$ and anisotropic $(\delta_2)$ pairing order parameters for the $E$ representation in Eq.\ref{e-3d}. }
    \label{fig:3bd}
\end{figure}

For the superconducting pairing in this  multi-orbital t-J model,  the intraorbital pairing is always dominant  over  the inter-orbital pairing as shown in previous work\cite{Seo-prl08pairing,yin2014spin}. Therefore, we can focus on the superconducting states with only intraorbital pairing. For the three $T_{2g}$ orbitals, the pairing operators between them  can be classified according to the irreducible representations of the  $T_d$ group as $\hat{T}_2\bigotimes \hat{T}_2=\hat{A}_1\bigoplus \hat{E}\bigoplus \hat{T}_1\bigoplus \hat{T}_2$, in which only $\hat{A}_1$ and $\hat{E}$ are formed by the intraorbital pairings. 
Combining with the momentum form factors in Eq.\ref{equ:pairingA}-\ref{equ:pairingT},  we can construct the BCS mean field decoupling terms:
\begin{align}
	H_{BCS}^A=&\Delta^{A}(\mathbf{k})\hat{A}_1+\delta_2 [ e(\mathbf{k})\hat{E}]_A,\label{equ:BCS-A}\\
	H_{BCS}^{E\pm}=&\Delta^{E_{\pm}}(\mathbf{k}) \hat{A}_1+\delta_2 [a(\mathbf{k})\hat{E}]_E+\delta_3 [e(\mathbf{k})\hat{E}]_E,
  \label{e-3d}\\
 	H_{BCS}^{T}=&\Delta^{T}(\mathbf{k}) \hat{A}_1+\delta_2[t_2(\mathbf{k})\hat{E}]_T,\label{equ:BCS-T}
\end{align}
with   the $\delta_{2,3}$ terms defined in App.\ref{app:pairing}
 The first terms in all above three pairing equations  represent  isotropic pairing among all three orbitals. As shown in the inset of Fig.\ref{fig:3bd}, in all  pairing states, this term is dominant over other pairing terms in our calculations.  Thus,  in the following paper,   we   simply focus on   the isotropic intraorbital pairings.
 
Under the slave boson mean-field approach, we study superconducting pairings in the hole doped  three-orbital model.  Due to the high degeneracy of the three $T_{2g}$ orbitals, the carrier occupancies in the three orbitals are identical, leading to an equal renormalization factor for all the hopping interactions\cite{gu-prb19Ni}.  The results are reported in Fig.\ref{fig:3bd}. Similar to the single-orbital model, the time reversal symmetry broken $d_{2z^2-x^2-y^2}\pm i d_{x^2-y^2}$ wave is  the most favored state.   In this calculation, we take the hopping parameters fitted to CoN and $J=0.2eV$.  Following the Hu-Ding principle\cite{hu_local_2012}, the t-wave pairing is expected to be weaker in the three-orbital model than in the one-orbital model because of  the band splitting at the $L$ points, which is  caused by the $t_3$ inter-orbital hopping as shown in Fig.\ref{fig:zb_disp}(b).   Our calculation confirms this result. Even in the heavy electron doped region, the t-wave never wins over the $d\pm id$ wave. Moreover, the splitting at $L$ points results in one Fermi surface along the $\Gamma$-$L$ lines. Thereafter, in the three-orbital model, eight gapless points appear in the $d\pm id$ wave state, resembling  the nodal d-wave state in cuprates.  The inset of Fig.\ref{fig:3bd}(a) shows that  the anisotropic pairing parameter  in the second term of the  $E$ representation in Eq.\ref{e-3d} quickly diminishes with increasing  hole doping  from half filling. The pairing prefers to maintain equal amplitudes on different orbitals as a result of the high degeneracy of the three $T_{2g}$ orbitals. This is consistent with the ``superconducting fitness" analysis for the multi-orbital pairing\cite{Sigrist-prb18fitness}, as the pairing prefers to take place among the electrons  on the same band with the same energy.

In summary, guided by the recent ideas on searching for potential new high temperature superconductors, we studied the superconducting states of a 3D t-J model in both one and three orbital cases under the slave boson meanfield approximation. It was found that the time reversal symmetry broken $d_{2z^2-x^2-y^2}\pm id_{x^2-y^2}$ wave superconducting state  wins over all other superconducting states upon hole doping. The state has gapless nodal points along the diagonal directions, being a direct3D extension of the d-wave superconducting state of cuprates.  It is also proper to mention that the  3D $d\pm id$ pairing symmetry was  theoretically suggested  previously for the $\beta$-pyrochlore CsW$_2$O$_6$\cite{Mazin-prb16} by a weak coupling approach\cite{Agterberg-prb99} without conclusive experimental evidence. 

To realize the physics studied in this paper, we must have a zinc-blende transition metal compound with a  $d^7$ filling configuration at transition metal atoms. As we have mentioned in the paper, CoN is a material  close to a realization of our model. However, as the Co atom in CoN has valence Co$^{3+}$, it is a $d^6$ filling configuration, namely it is  heavily hole doped material with respect to a $T_{2g}$ half filled $d^7$ configuration. Thus, we may consider CoO with a zinc-blende structure to be the parental compound since  the Co atom has valence Co$^{2+}$  with a $T_{2g}$ half filled $d^7$ configuration.  The zinc-blende CoO was known to be a metastable\cite{CoO62} and have an AFM ground state\cite{PR58-fccMag,CoO10}. The CoN was experimentally known to be a very good metal\cite{CoNO15}.   Therefore, assuming that the zinc-blende structure is not drastically modified locally,  we can suggest that the potential superconductivity may be realized in CoO$_{1-x}$N$_x$\cite{CoNO04} in which x describes the concentration of  doped hole carriers. Under the same assumption,  we may also consider materials by adding electron carriers to CoN, such as (H, Li)$_{1-x}$CoN\cite{reddy2013li,MPLB18-1rsprinciple}, to realize potential superconducting states.

 The model has very high symmetry in both orbital and lattice spaces.  Here we only consider the superconducting states with respect to the full lattice symmetry.  In principle,
 the degeneracy among three orbitals and the cubic lattice symmetry can be broken spontaneously in many different ways by a variety of mechanisms. Therefore, it can be extremely interesting to explore novel electronic states in future.

{\it Acknowledgement:} Q. Zhang acknowledges the support from the International Young Scientist Fellowship of Institute of Physics CAS (Grant No. 2017002) and the Postdoctoral International Program from China Postdoctoral Science Foundation (Grant No. Y8BK131T61). Y. Gu is supported by the High-performance Computing Platform of Peking University. JP. Hu is supported by the Ministry of Science and Technology of China 973 program(Grant No. 2015CB921300, No.~2017YFA0303100), National Science Foundation of China (Grant No. NSFC-11334012), and   the Strategic Priority Research Program of  CAS (Grant No. XDB07000000).


\section{Landau Theory for Pairing Symmetries}\label{app:landau}
The pairing form can be a linear combination of the basis function of a representation $\Delta(\mathbf{k})=\sum_i \delta_i f_i(\mathbf{k})$ with $f_i(\mathbf{k})$ being the basis functions and  $\delta_i$ being the pairing amplitude. In our consideration, the electron pairing takes place on the NN bond mediated by AFM superexchange. Considering an equal on-bond pairing amplitude, as mentioned in the main text, the pairing belongs to the $E$ representation is fixed as $\delta_d(e_1 \pm i e_2 )$. However, for the $T_2$ representation,  $\Delta^T =\delta_x s_zs_y+\delta_y s_xs_z+\delta_z s_xs_y$, there is additional phase freedom among the $\delta_i$'s. 

We now use Ginzburg-Landau theory to explore the phase freedom of $T_2$ representation. The free energy respects all the symmetries including point group, time reversal, $U(1)$ gauge symmetry and spin rotational $SU(2)$ symmetry\cite{Sigrist-RMP91}. The superconducting phases are determined by  the fourth order terms.  We extract the invariant $A_1$ forms from $T_2\otimes T_2^\ast\otimes T_2\otimes T_2^\ast$. They are
\begin{itemize}
    \item $(|\delta_x|^2+|\delta_y|^2+|\delta_z|^2)^2$,
    \item $|\delta_x|^4+|\delta_z|^4+|\delta_z|^4-(|\delta_y\delta_z|^2+|\delta_x\delta_z|^2+|\delta_y\delta_x|^2)$,
    \item $(\delta_x\delta^\ast_y\pm \delta^\ast_x\delta_y)^2+(\delta_x\delta^\ast_z\pm \delta^\ast_x\delta_z)^2+(\delta_z\delta^\ast_y\pm \delta^\ast_x\delta_z)^2$.
\end{itemize}

Those are essentially the polynomial invariants as we project $|\delta_x+\delta_y+\delta_z|^4$ on the basis functions. Only three of them are independent and they can be rephrased with three independent parameters in the fourth order term:
\begin{align}
\Delta F_4(\bm{\delta})=&\beta_1(|\delta_x|^2+|\delta_y|^2+|\delta_z|^2)^2+\beta_2(|\delta_y\delta_z|^2+|\delta_x\delta_z|^2\nonumber\\
			&+|\delta_y\delta_x|^2)+\beta_3|\delta^2_x+\delta^2_y+\delta^2_z|^2.
\end{align}
Considering the equal NN bonds pairing amplitude $\bm{\delta}=\delta_t(1,e^{i\theta_1},e^{i\theta_2})$, mediated by the NN AFM exchange interactions,  the free energy becomes 
\begin{align}
   \Delta F_4(\bm{\delta})= \delta^4_t(9\beta_1+3\beta_2+\beta_3|1+e^{i2\theta_1}+e^{i2\theta_2}|^2). 
\end{align}
The minimization of $\Delta F_4(\bm{\delta})$ gives the following phases characterized by $(\theta_1,\theta_2)$:
\begin{itemize}
    \item TRI states: $\beta_3<0$ and $3\beta_1+\beta_2+3\beta_3>0$
        \begin{itemize}
            \item ferro: $(0,0)$
            \item antiferro: $(0,\pi)$ or $(\pi,0)$ or $(\pi,\pi)$;
        \end{itemize}
    \item Chiral states:  $\beta_3>0$ and $3\beta_1+\beta_2>0$
        \begin{itemize}
            \item $(4\pi,2\pi)/3$ or $(2\pi,4\pi)/3$
            \item $(4\pi,5\pi)/3$ or $(5\pi,4\pi)/3$
            \item $(\pi,2\pi)/3$ or $(2\pi,\pi)/3$
            \item $(\pi,5\pi)/3$ or $(5\pi,\pi)/3$.
        \end{itemize}
\end{itemize}
The TRI state $s_zs_y\pm s_xs_z\pm s_xs_y$ are fourfold degenerate and the chiral state  $s_zs_y\pm \omega s_xs_z\pm \omega^\ast s_xs_y$ state are eight-fold degenerate with $\omega=e^{\pm i2\pi/3}$. The states marked with $T_{20}$ and $T_{21}$ are the states $(0,0)$ and $(4\pi,2\pi)/3$, respectively.

\section{Symmetry Analysis for Three-orbital Tight-binding Model}\label{app:symmetry}
For the three-orbital tight-binding model, the Hamiltonian  $H=\sum d^\dagger_{\alpha\mathbf{k}} H_{3t,\alpha\beta}(\mathbf{k}) d_{\beta\mathbf{k}}$ shall be the $A_1$ representation of the point group. It contains the orbital parts and the momentum factor parts. The orbital parts are the representation product and it can be decomposed as : $\hat{T}_2\bigotimes \hat{T}_2=\hat{A}_1\bigoplus \hat{E}\bigoplus \hat{T}_1\bigoplus \hat{T}_2$. To make the Hamiltonian $A_1$ representation, the factor parts should  be the same  representations with the orbital parts. Firstly considering the intra-orbital hopping terms, the orbital parts can form $\hat{A}_1$ representation as $d^\dagger_{yz}d_{yz}+d^\dagger_{xz}d_{xz}+d^\dagger_{xy}d_{xy}$ and $\hat{E}$ representation as $(2d^\dagger_{xy}d_{xy}-d^\dagger_{yz}d_{yz}-d^\dagger_{xz}d_{xz}, d^\dagger_{yz}d_{yz}-d^\dagger_{xz}d_{xz})$. Combining with the factor of $A_1$ and $E$ forms $(e_1 ,e_2 )$, they can be rephrased as the $t_1$ and $t_2$ terms in the Hamiltonian. For the inter-orbital hopping, the $\hat{T}_2$ representation basis can be written as $(d^\dagger_{xz}d_{xy},d^\dagger_{yz}d_{xy},d^\dagger_{yz}d_{xz})+h.c.$, with $h.c.$ denoting the Hermitian conjugate counterparts. Combining with the factors $(s_ys_z,s_xs_z,s_xs_y)$, they are the $t_3$ terms. The above terms do not break inversion symmetry so that $t_{1,2,3}$ are real numbers. Due to the existence of nitrogen atoms, the $O_h$ group reduces to $T_d$ for the breach of inversion symmetry. This allows us to write down the $t_4$ term which breaks inversion symmetry. They come from the combination of $\hat{T}_1$ representation of the orbitals $i(d^\dagger_{xz}d_{xy},d^\dagger_{yz}d_{xy},d^\dagger_{yz}d_{xz})+h.c$ and the factors $(s_x(c_y-c_z),s_y(c_z-c_x),s_z(c_x-c_y))$.   From time reversal symmetry and Hermitian condition of the Hamiltonian, $t_4$ is  real.

\section{Three-orbital Model Pairing Forms}\label{app:pairing}
In   the pairing part of Eq.(\ref{equ:BCS-A}-\ref{equ:BCS-T}) of the three-orbital model, the orbital quadratic forms $\hat{A}_1$ and $\hat{E}$ are:
\begin{align}
    \hat{A}_1=&d_{yz}^\dagger d_{yz}^\dagger +d_{xz}^\dagger d_{xz}^\dagger +d_{xy}^\dagger d_{xy}^\dagger ,\nonumber\\
    \hat{E}^1=&2d_{xy}^\dagger d_{xy}^\dagger -d_{xz}^\dagger d_{xz}^\dagger -d_{yz}^\dagger d_{yz}^\dagger,\\
    \hat{E}^2=&\sqrt{3}(d_{yz}^\dagger d_{yz}^\dagger-d_{xz}^\dagger d_{xz}^\dagger ), \nonumber
\end{align}
where the spin and momentum labels are omitted for the momentum conserved spin-singlet pairing forms.  The additional pairing forms of $\delta_{2,3}$ are 
\begin{align}
[e(\mathbf{k})\hat{E}]_A=&\sum_{\sigma=\pm}\Delta^{E_{\sigma}}(\mathbf{k})(\hat{E}^1-i\sigma\hat{E}^2),\\
[a(\mathbf{k})\hat{E}]_E=&4(c_xc_y+c_yc_z+c_xc_z)(\hat{E}^1\pm i\hat{E}^2),\\
[e(\mathbf{k})\hat{E}]_E=& (e_1 \hat{E}^1-e_2 \hat{E}^2\pm i(e_1 \hat{E}^2+e_2 \hat{E}^1)),\\
[t_2(\mathbf{k})\hat{E}]_T=&s_ys_z(\hat{E}^1-\sqrt{3}\hat{E}^2)+e^{i\theta_1}s_xs_z(\hat{E}^1+\sqrt{3}\hat{E}^2)\nonumber\\
						&+e^{i\theta_2}2s_xs_y\hat{E}^1,
\end{align}
with   $e_1 =2c_xc_y-c_xc_z-c_yc_z$ and $e_2 =\sqrt{3}(c_yc_z-c_xc_z)$.

%


\end{document}